**Optimized Nanogap Thermophotovoltaic Devices for Waste Heat Recovery**

Mehran Habibzadeh[a] and Sheila Edalatpour[a,b*]

[a]*Department of Mechanical Engineering, University of Maine, Orono, Maine 04469, USA*

[b]*Frontier Institute for Research in Sensor Technologies, University of Maine, Orono, Maine 04469, USA*

*Email: sheila.edalatpour@maine.edu

## Abstract

Nanogap thermophotovoltaic (TPV) devices can deliver high power densities even with the medium-temperature heat sources. As such, these devices are very promising for recovering industrial waste heat. So far, the demonstrated nanogap TPVs have shown performances far below optimal. The objective of this study is to identify the optimal designs for nanogap TPV devices targeted for industrial waste heat recovery. Optimal configurations for maximal power density, maximal efficiency, and a trade-off between the two are determined as a function of the size of the vacuum gap between the emitter and the photovoltaic (PV) cell. The effects of adding a metallic cover to the PV cell, as well as introducing an air gap between the PV cell and the reflector, are also studied through this optimization framework. Results show that the optimal device configuration is highly sensitive to the vacuum gap size. A metallic cover enhances power density for gaps below 125 nm due to surface plasmon-polariton coupling, but significantly reduces efficiency due to its parasitic absorption. To realize the benefits of air gaps, ultrathin PV cells requiring mechanical support by a substrate are needed. The presence of the substrate, however, diminishes the benefits of the air gap rendering them ineffective. ITO and InAs are found as optimal materials for the emitter and PV cell, respectively, owing to ITO's tunable plasma frequency and InAs's low bandgap.

**1. Introduction**

Thermophotovoltaic (TPV) devices are solid-state heat engines that convert thermal energy directly into electricity [1]. TPV systems are promising heat engines due to their numerous advantages, including simplicity, low maintenance requirements, compatibility with a wide range of heat sources, and silent operation [1,2]. Recent experimental studies demonstrating record efficiencies of 40% [3] and 44% [4], achieved with emitter temperatures of 2400 °C and 1435 °C, respectively, have sparked great interest toward TPV technology. Currently, one of the main applications of TPVs is in thermal energy storage systems where they are employed for efficient recovery of stored heat as electricity [5-10].

The simplicity and flexibility of TPV systems make them a promising candidate for recovering industrial waste heat [5]. However, their practical deployment has been limited by relatively low efficiency and power density when operating with typical industrial waste heat. These heat sources have much lower temperatures than the emitters used in the thermal energy storage systems [11-13]. Nanogap TPVs which utilize a nanoscale separation distance between the emitter and the photovoltaic (PV) cell, have exhibited substantial improvements in both efficiency and power density compared to conventional far-field TPVs operating at the same emitter temperature [14-16]. This performance enhancement in nanogap TPVs arises from additional radiative heat transfer between the emitter and PV cell, mediated by evanescent waves generated through frustrated total internal reflection (FTIR) and surface modes. To date, nanogap TPVs have been experimentally demonstrated in nine studies [17-25]. These experimental studies represented power enhancements ranging from 1.1 to 40 times compared to the far-field performance of the same devices using emitter temperatures between 382 °C and 997 °C. These studies have significantly progressed the field by providing proof-of-concept demonstrations for nanogap TPVs as solid-state heat engines. However, most experimentally-demonstrated nanogap TPVs operate well below their optimal performance. Indeed, the optimal design of nanogap TPVs for waste heat recovery is not well understood. There is a demand for nanogap TPV designs which render both high power density and efficiency at industrial waste heat temperatures, while also being scalable and practical. The current study aims to address this need by performing an optimization analysis to identify nanogap TPV designs best suited for waste heat recovery applications. The optimization is performed to achieve three objectives, namely maximal power density, maximal efficiency, and an optimal trade-off between the two. These objectives are each evaluated across eight vacuum gap sizes, ranging from 25 to 200 nm. The optimization is conducted by using a genetic algorithm and deep learning. State-of-the-art enhancement strategies are applied into the analysis, including the use of a back reflector for the PV cell [26-33], the deposition of a thin metallic film on the PV cell [34,35], and the implementation of an air gap between the PV cell and the back reflector [4,36-38]. In addition, practical constraints are considered to ensure the proposed optimal designs are suitable for real-world applications. These constraints include using a supporting buffer layer for thin free-standing PV cells, selecting buffer materials with matching thermal expansion coefficients, and considering a minimum thickness of 3 nm for the metallic cover to account for fabrication limitations. The optimal designs for various gap sizes are determined and the physics underlying the optimal designs of these devices are discussed. This study advances the theoretical understanding of nanogap TPV systems and can guide the development of high performance and practical nanogap TPVs for industrial waste heat recovery.

The remainder of this paper is organized as follows. Section 2 describes the detailed balance method used to model the performance of nanogap TPVs, as well as the utilized optimization technique based on genetic algorithms and deep learning. The problem under study is described in Section 3, and the results are presented and discussed in Section 4. Concluding remarks are provided in Section 5. The material properties used in the simulations, along with the parameters of the optimal designs as a function of gap size, are provided in supplementary materials.

## 2. Methods

### 2.1. Theoretical models for predicting power density and efficiency

A schematic of the considered nanogap TPV device is shown in Fig. 1. A thermal emitter (medium 1) is separated from a PV cell (medium 4) via a vacuum gap (medium 2). The PV cell is covered with a thin metallic cover (medium 3), while separated from a back reflector (medium 7) via an air gap (medium 6). A buffer layer (medium 5) is considered to provide mechanical support for the PV cell when its thickness is insufficient to be self-supporting. The emitter and the back reflector are considered to be optically thick, while the thicknesses of the vacuum gap, metallic cover, PV cell, buffer and air gap are denoted by $d$, $t_{\mathrm{m}}$, $t_{\mathrm{PV}}$, $t_{\mathrm{b}}$ and $t_{\mathrm{a}}$, respectively. The materials comprising the emitter, metallic cover, PV cell, buffer, and back reflector are represented by $M_{\mathrm{e}}$, $M_{\mathrm{m}}$, $M_{\mathrm{PV}}$, $M_{\mathrm{b}}$, and $M_{\mathrm{br}}$, respectively. The temperature of medium $j$ ($j$ = 1, 2, …, 7) is $T_j$, and all media are assumed to be infinite along the $x$- and $y$-axes (the coordinate system is shown in Fig. 1). The objective is to find the maximum power density, $P$, and efficiency, $\eta$, of the TPV device.

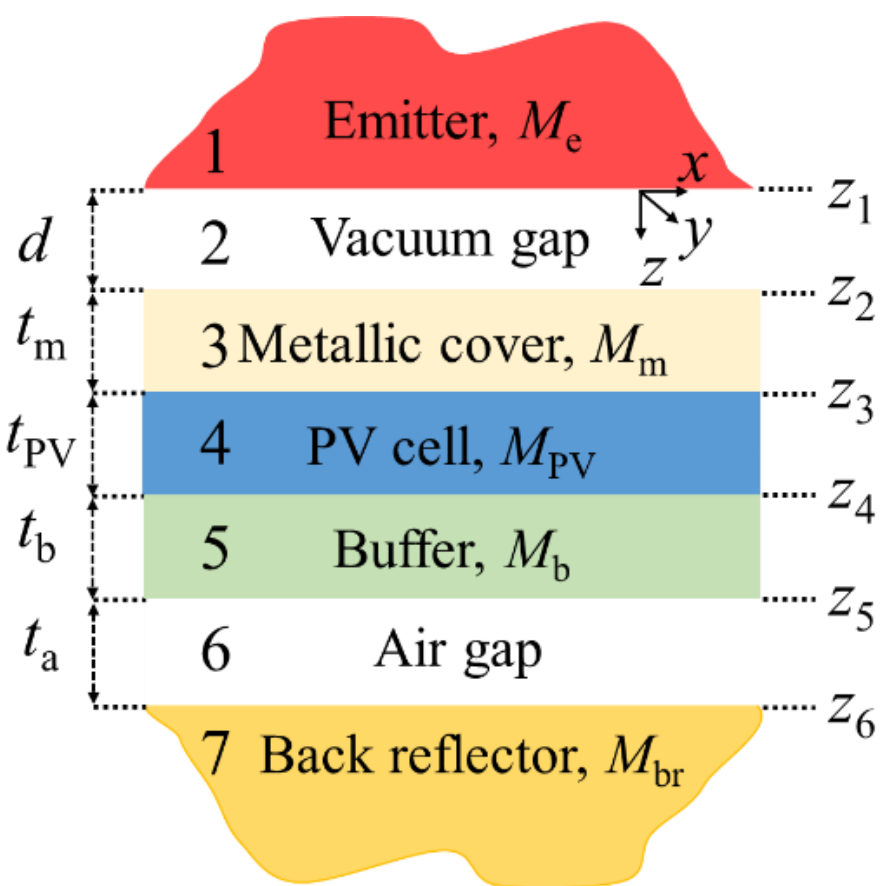

Fig. 1. A schematic of a nanogap TPV device featuring a metallic cover on the PV cell and an air gap between the buffer layer and the back reflector.

The maximum power density, referred to as power density for simplicity hereafter, can be found as $P = \max(I.V)$, where $V$ is the applied voltage and $I$ is the photocurrent density defined as the photocurrent generated by the PV cell per unit surface area. The photocurrent density is given by [39]:

$$I = e\left[G - R_{\mathrm{rad}}(V) - R_{\mathrm{Auger}}(V) - R_{\mathrm{SRH}}(V)\right] \tag{1}$$

where, $e$ is the elementary charge, $G$ is the rate of electron-hole pair (EHP) generation in the PV cell, and $R_{\mathrm{rad}}(V)$, $R_{\mathrm{Auger}}(V)$, and $R_{\mathrm{SRH}}(V)$ show the rates of radiative, Auger, and Shockley-Read-Hall (SRH) recombination of the EHPs, respectively. Electron-hole pairs are generated in the PV cell when photons with energy greater than the bandgap energy of the PV cell are thermally radiated from the emitter and are absorbed by the PV cell. The rate of above-bandgap photons absorbed in the PV cell can be found as:

$$G = \int_{\omega_g}^{\infty} G_{\omega,14} d\omega \tag{2}$$

where $\omega$ is the angular frequency, $\omega_g$ is the angular frequency associated with the bandgap energy of the PV cell, and $G_{\omega,14}$ shows the rate of absorption of photons with a given frequency $\omega$ in the PV cell (medium 4) due to thermal radiation from the emitter (medium 1). The rate of photon absorption in medium $j$ due to thermal emission by medium $i$ can be calculated as:

$$G_{\omega,ij} = F_{\omega,ij}(z_{j-1}) - F_{\omega,ij}(z_j) \quad (3)$$

where, as shown in Fig. 1, $z_{j-1}$ and $z_j$ denote the $z$ coordinates of the top and bottom interfaces of medium $j$, respectively. The term $F_{\omega,ij}(z_l)$ is the flux of photons with frequency $\omega$ crossing surface $z = z_l$ in medium $j$ due to thermal radiation from medium $i$ and can be found:

$$F_{\omega,ij}(z_l) = \int_0^\infty F_{\omega,k_\rho,ij}(z_l) dk_\rho \quad (4)$$

The term $F_{\omega,k_\rho,ij}(z_l)$ in Eq. 4 is the photon flux associated with a photon with momentum $k_\rho$, which can be obtained using fluctuational electrodynamics as [40]:

$$F_{\omega,k_\rho,ij}(z_l) = \frac{k_0^2}{\pi^2\left[\exp\left(\frac{\hbar\omega - eV_i}{k_B T_i}\right)-1\right]} \mathrm{Re}\left\{\mathrm{i}\varepsilon_i''(\omega) k_\rho \int_{z_{i-1}}^{z_i} dz' \left[g_{ij,\rho\rho}^{E}(k_\rho, z_l, z', \omega) g_{ij,\theta\rho}^{H*}(k_\rho, z_l, z', \omega) + g_{ij,\rho z}^{E}(k_\rho, z_l, z', \omega) g_{ij,\theta\rho}^{H*}(k_\rho, z_l, z', \omega) - g_{ij,\theta\theta}^{E}(k_\rho, z_l, z', \omega) g_{ij,\rho\theta}^{H*}(k_\rho, z_l, z', \omega)\right]\right\} \quad (5)$$

In Eq. 5, $k_0$ is the magnitude of the wavevector in the free space, $\hbar$ and $k_B$ are the reduced Planck and Boltzmann constants, respectively, $V_i$ is the voltage applied to medium $i$ ($V_i = 0$ when $i \neq 4$), Re and subscript * indicate the real part and the complex conjugate of a complex number, respectively, i is the imaginary unit, $\varepsilon_i''$ is the imaginary part of the dielectric function of the medium $i$, $k_\rho$ is the parallel component of the wavevector (or the momentum of photons), $\rho$ is the radial coordinate, $\theta$ is the azimuthal angle, and $g_{ij,\alpha\beta}^{E}$ and $g_{ij,\alpha\beta}^{H}$ are the $\alpha\beta$ ($\alpha$ and $\beta$ = $\rho$, $\theta$, and $z$) components of the Weyl expansion of the electric and magnetic dyadic Green's functions between layers $i$ and $j$, respectively. The Weyl components of the dyadic Green's functions can be found using the scattering matrix method as detailed in Ref. [40]. The rate of radiative recombination in the PV cell can be found as the rate of photons with an energy greater than $\omega_g$ radiated from the PV cell and absorbed by the emitter, i.e.,

$$R_{\text{rad}}(V) = \int_{\omega_g}^{\infty} G_{\omega,41} d\omega \tag{6}$$

where $G_{\omega,41}$ can be found using Eq. 3.

The Auger recombination rate can be found as [1]:

$$R_{\text{Auger}}(V) = (C_e n + C_h p)(np - n_i{}^2) t_{\text{PV}} \tag{7}$$

where $C_e$ ($C_h$) is the Auger recombination coefficients for electrons (holes), $n_i$ is the intrinsic carrier concentration in the PV cell, and $n$ and $p$ are electron and hole concentrations in the PV cell, respectively. The electron and hole concentrations can be found as [1]:

$$n = N_c \exp\left(\frac{E_{F,n} - E_C}{k_B T_4}\right) \tag{8a}$$

$$p = N_v \exp\left(\frac{E_v - E_{F,p}}{k_B T_4}\right) \tag{8b}$$

where $N_{\text{c}}$ ($N_{\text{v}}$) is the effective density of states in the conduction (valence) band, $E_c$ ($E_v$) is the conduction (valence) band edges, and $E_{F,n}$ and $E_{F,p}$ are the quasi-Fermi level of electrons and holes, respectively, which can be obtained as [1]:

$$E_{F,n} = \frac{k_B T_4}{2}(-\ln N_{\text{c}} + \ln N_{\text{v}} - \ln N_{\text{A}} + \ln N_{\text{D}}) + \frac{1}{2}(E_{\text{c}} + E_{\text{v}} + eV) \tag{9a}$$

$$E_{F,p} = E_{F,n} - eV \tag{9b}$$

In Eqs. 9a and 9b, $N_{\text{A}}$ and $N_{\text{D}}$ are the acceptor and donor concentrations, respectively.

The SRH recombination rate is given by [1]:

$$R_{\mathrm{SRH}}(V) = \frac{(np-n_i{}^2)t_{\mathrm{PV}}}{\tau(n+p+2n_i)} \tag{10}$$

where $\tau$ is the SRH lifetimes of the charge carriers.

The efficiency, $\eta$, of the TPV device is defined as:

$$\eta = \frac{P}{P_{rad}} \tag{11}$$

where $P_{rad}$ is the net radiative heat flux transferred across the vacuum gap found using the rate of photons radiated from the emitter and absorbed by media 3 to 7 across the vacuum gap:

$$P_{rad} = \int_0^{\infty} \hbar\omega \left[\sum_{j=3}^{7}\left(G_{\omega,1j} - G_{\omega,j1}\right)\right] d\omega \tag{12}$$

where $G_{\omega,1j}$ and $G_{\omega,j1}$ can be calculated using Eq. 3.

### 2.2. Deep learning-assisted optimization

A multi-objective genetic algorithm is utilized to optimize the structure of the nanogap TPV device. The design variables include the thicknesses and materials of the various layers utilized in the device (see Fig. 1), while the objective functions are the power density and efficiency. The MATLAB's built-in gamultiobj function has been used for genetic algorithm-based optimization. A schematic of the optimization process is shown in Fig. 2a. The optimization process starts with a randomly generated initial population. The power density and efficiency are predicted for each individual in the population, and the individual's fitness is evaluated using a fitness function. If the fitness values do not satisfy predefined criteria, a new population is generated through crossover and mutation operations applied to the best-performing individuals. This cycle is continued over multiple generations until the stopping criteria are satisfied.

To decrease the computational cost of evaluating power density and efficiency, a deep learning model is trained for each considered vacuum gap using the data generated by the theoretical models explained in Section 2.1. These deep-learning models replace the direct use of theoretical models, which substantially accelerate the optimization process. Consequently, larger population sizes and multiple optimization runs with different hyperparameters become possible, which increases the possibility of identifying a globally optimal design. The deep learning models are implemented in MATLAB using the built-in fullyConnectedLayer function. As illustrated in Fig. 2, the models are designed as multi-layer perceptron neural networks with two to three hidden layers composed of fully connected layers.

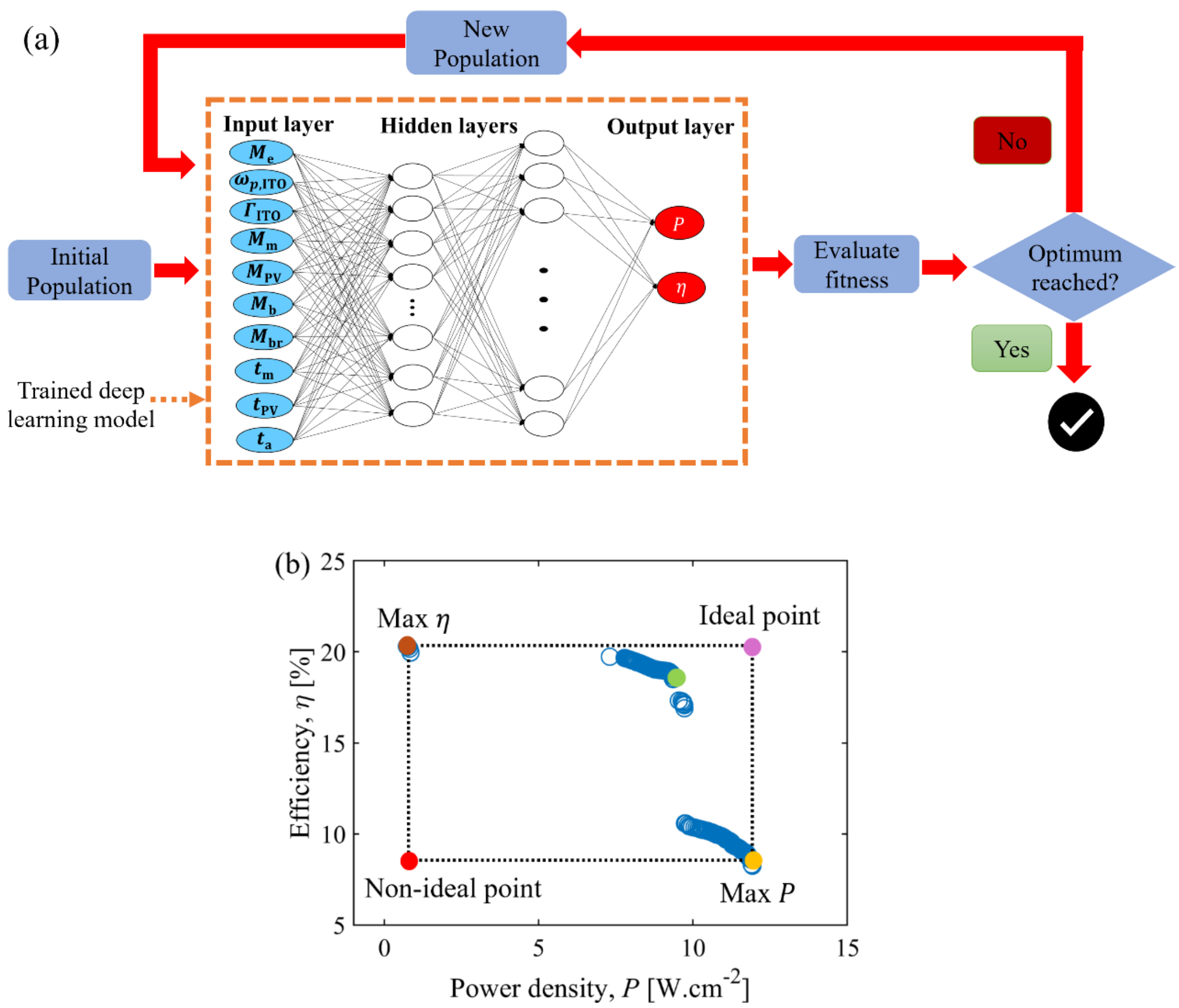

Fig. 2. (a) Schematic representation of the optimization process using a genetic algorithm integrated with deep learning. (b) Pareto frontier plot for the MC configuration and a vacuum gap size of $d$ = 25 nm.

**3. Problem under study**

The system under study is schematically shown in Fig. 1. A temperature of 900 K is assumed for the emitter while the layers located across the vacuum gap are considered at 300 K. The optimization is done for eight vacuum gaps ranging from 25 to 200 nm. The design variables and their corresponding domains are provided in Table 1.

The design variables include the materials selected for the emitter, metallic cover, PV cell, buffer layer, and back reflector, in addition to the thicknesses of the metallic film, PV cell, and air gap. Four materials, namely indium tin oxide (ITO), tungsten (W), graphite (G), and silicon carbide (SiC), are selected for the emitter because of their high thermal stability and common use in TPV devices [41,42]. It should be noted that the plasma frequency ($\omega_{\mathrm{p,ITO}}$) and damping rate ($\Gamma_{\mathrm{ITO}}$) of ITO can be tuned by adjusting the oxygen content during fabrication [34], and thus have been selected as design variables. Seven metals, namely platinum (Pt), gold (Au), silver (Ag), aluminum (Al), titanium (Ti), cobalt (Co), and copper (Cu) are considered as the metallic cover material. The thickness of the metallic film is allowed to vary between 3 and 20 nm. While ultra-thin metal films with a thickness of only 2 nm have been experimentally demonstrated [43,44], a lower limit of 3 nm is fixed here to ensure fabrication practicality. It should be noted that the metallic cover cannot be used as the primary layer for electron extraction and lateral current spreading due to its very small thickness, which would result in large ohmic losses. However, electron extraction can be done via a backside electrode, while the metallic layer can serve as a counter-contact that closes the external circuit. Indium arsenide (InAs) with a low bandgap of 0.354 eV and gallium antimonide (GaSb) with a moderate bandgap of 0.726 eV are considered for the PV cell. The thickness of the PV cell is allowed to vary from 10 to 5000 nm. For ultra-thin PV cells, a buffer layer is necessary to prevent buckling in the presence of an air gap. Assuming the air gap is maintained using micropillars deposited on the back reflector, PV cells with a thickness less than 5% of the micropillar spacing require mechanical support from the buffer layer [45]. With a micropillar spacing of 10 μm, PV cells thinner than 500 nm must be supported. To ensure sufficient mechanical stability, the buffer layer thickness is fixed at 500 nm. The buffer layer material is selected from gallium phosphide (GaP), gallium arsenide (GaAs), and indium phosphide (InP), which have bandgaps of approximately 2.2, 1.4, and 1.3 eV, respectively. These materials are chosen mainly due to their large bandgaps, which minimize parasitic optical absorption. Additionally, these materials can be fabricated and doped with low defect density using well-established fabrication techniques [46-47]. The N-type doped variants of these materials are considered, as they enhance back-side electron collection and minimally interfere charge transport [48]. The air gap thickness can vary from 500 nm to 5000 nm, and the same materials used for the metallic cover are also considered for the back reflector. The dielectric function and the parameters used for modeling the non-radiative recombination rates using Eq. 7, 9 and 10 are presented in part A of the supplementary materials.

Table 1. Design variables and their domain.

| **Variable** | **Domain** |
|---|---|
| $M_e$ | ITO, SiC, W, Graphite |
| $\omega_{p,\mathrm{ITO}}$ | 0.4 – 0.9 eV |
| $\Gamma_{\mathrm{ITO}}$ | 0.1 – 0.15 eV |
| $M_{\mathrm{m}}$ and $M_{\mathrm{br}}$ | Pt, Au, Ag, Al, Ti, Co, Cu |
| $M_{\mathrm{PV}}$ | InAs, GaSb |
| $M_{\mathrm{b}}$ | GaP, GaAs, InP |
| $t_{\mathrm{m}}$ | 3 – 20 nm |
| $t_{\mathrm{PV}}$ | 10 – 5000 nm |
| $t_{\mathrm{a}}$ | 500 – 5000 nm |

In addition to the TPV system incorporating both a metallic cover and an air gap (MC+AG), as shown in Fig. 1, three other configurations are examined: one with only a metallic cover (MC), one with only an air gap (AG), and one with neither (Basic). This comparative analysis is performed to identify the optimal configurations for the nanogap TPV system.

## 4. Results and discussion

A deep learning model is trained for each of the four configurations under study (i.e., Basic, AG, MC, and MC+AG) and for each of the eight considered vacuum gap sizes. For every configuration and gap size, 50,000 samples are generated within the defined design domain presented in Table 1 using the theoretical model described in Section 2.1. Of the generated data, 70% is used for training the deep learning model, while the remaining 30% is equally split between validation and testing. The trained deep learning models have a coefficient of determination, $R^2$, greater than 0.99 and a normalized root-mean-square error (NRMSE) of less than 1% on the test data, demonstrating their high fidelity. Additionally, using the trained machine learning models reduces the total simulation time required for optimization by approximately 75%. For example, optimization based on analytical equations can take a few weeks on a single processing unit, while it can be completed in a few days when machine learning models are employed. This reduction mainly originates from the significant decrease in evaluation time per design, from tens of seconds with analytical equations to approximately instantaneous predictions with machine learning. It should be noted that, when using machine learning models, most of the computational cost is associated with generating the training and testing datasets using analytical equations. However, the size of these datasets is significantly smaller than the number of function evaluations required for genetic algorithm optimization, resulting in an overall reduction of 75% in computational cost.

The multi-objective optimization is performed to identify the optimal design variables for each considered configuration and gap size. At the end of the optimization process, a Pareto frontier plot, such as the one shown in Fig. 2b, is generated. The plot in Fig. 2b corresponds to the MC configuration with a gap size of 25 nm. It illustrates the non-dominated solutions, marked by blue circles, which represent the maximum achievable efficiency for a given power density. The non-dominated solution at the top-left corner, shown in brown, presents the maximum efficiency, which also corresponds to the lowest power density. In contrast, the non-dominated solution at the bottom-right corner, presented in yellow, shows the highest power density but the lowest efficiency. In addition, an ideal reference point highlighted in magenta and a non-ideal reference point highlighted in red are shown. These reference points are benchmarks rather than actual optimization results. The ideal point has the maximum possible values for both objectives, while the non-ideal point has their minimum values.

The Pareto plot is employed to identify optimal designs corresponding to maximum power density, maximum efficiency, and a balanced trade-off between the two. The trade-off solution, which is shown by a green point on Fig. 2b, is selected using the technique for order of preference by similarity to ideal solution (TOPSIS) [49]. In this technique, a closeness coefficient is computed for each non-dominated solution as the ratio of its distance to the non-ideal point to the sum of its distances to the ideal and non-ideal points. The non-dominated solution with the highest closeness coefficient is identified as the optimal trade-off point. The optimal designs corresponding to maximum power density, maximum efficiency, and the trade-off point are presented and discussed in the following sub-sections.

### 4.1. Optimal designs for maximum power density

Figure 3a shows the maximum power density found for each of the four configurations as a function of gap size, with the corresponding efficiencies shown in Fig. 3b. For each configuration, the design parameters that optimize power, in addition to the resulting power density and efficiency, are listed in Table S-3 in part B of the supplementary materials. Additionally, the configuration yielding the highest power density at each gap size is determined and listed in Table 2. The design parameters, power density, and efficiency of the optimal configurations are also presented in Table 2. Several important observations can be made from the obtained results. Configurations utilizing a metal cover have the highest power densities for gap sizes smaller than 125 nm. However, the effect of metal cover on the power density diminishes as the gap size increases. Moreover, the metal cover has a negative effect on efficiency for all gap sizes.

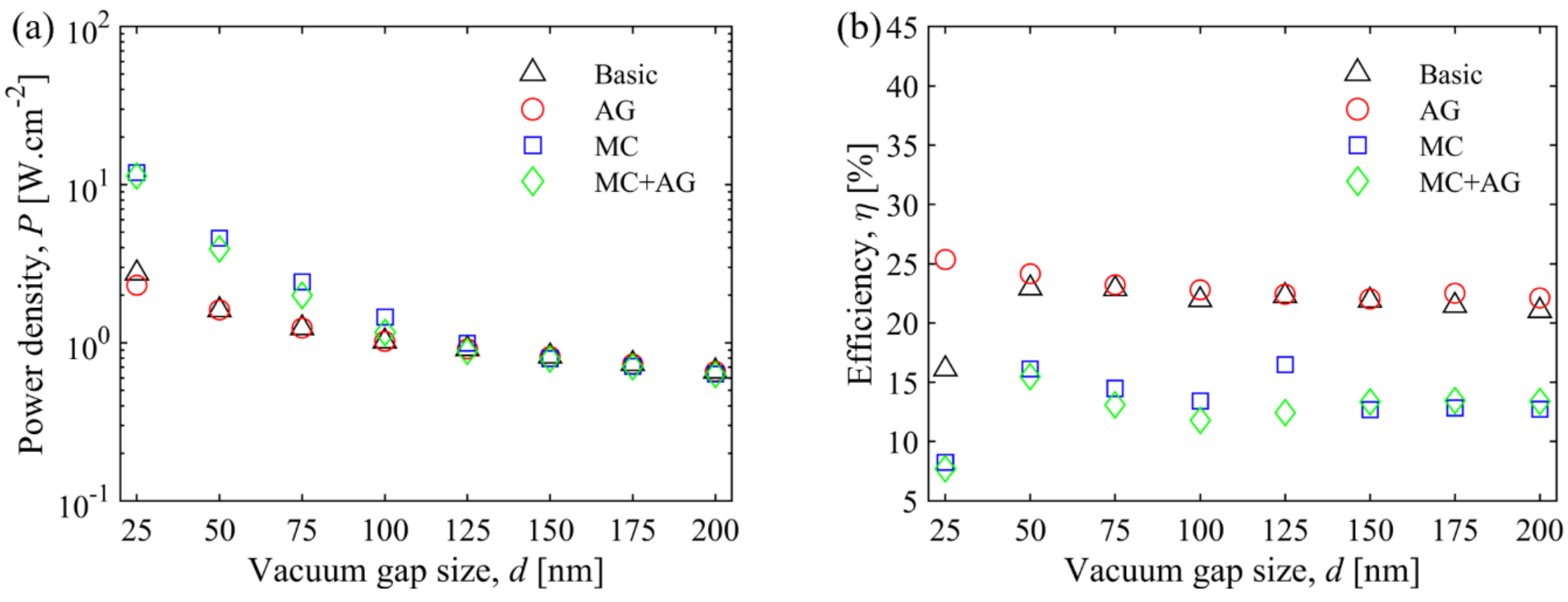

Fig. 3. (a) Power density and (b) efficiency of nanogap TPV devices optimized for maximal power density as a function of vacuum gap size.

Table 2. Optimal design variables and performance parameters for maximal *power density*.

| $d$ [nm] | Config. | $M_e$ | $\omega_{p,\mathrm{ITO}}$ [eV] | $\Gamma_{\mathrm{ITO}}$ [eV] | $M_m$ | $t_m$ [nm] | $M_{PV}$ | $t_{PV}$ [nm] | $M_b$ | $t_a$ [nm] | $M_{br}$ | $P$ [Wcm$^{-2}$] | $\eta$ [%] |
|---|---|---|---|---|---|---|---|---|---|---|---|---|---|
| 25 | MC | ITO | 0.51 | 0.1 | Co | 4 | InAs | 195 | - | 0 | Ti | 11.92 | 8.3 |
| 50 | MC | ITO | 0.53 | 0.1 | Cu | 3 | InAs | 255 | - | 0 | Ti | 4.57 | 16.1 |
| 75 | MC | ITO | 0.50 | 0.1 | Cu | 3 | InAs | 250 | - | 0 | Ti | 2.42 | 14.5 |
| 100 | MC | ITO | 0.49 | 0.1 | Cu | 3 | InAs | 240 | - | 0 | Ti | 1.46 | 13.4 |

| 125 | MC | ITO | 0.50 | 0.1 | Cu | 3 | InAs | 315 | - | 0 | Cu | 0.99 | 16.5 |
|---|---|---|---|---|---|---|---|---|---|---|---|---|---|
| 150 | Basic | SiC | - | - | - | 0 | InAs | 2185 | - | 0 | Cu | 0.82 | 21.9 |
| 175 | Basic | SiC | - | - | - | 0 | InAs | 2185 | - | 0 | Cu | 0.73 | 21.5 |
| 200 | Basic | SiC | - | - | - | 0 | InAs | 2180 | - | 0 | Cu | 0.66 | 21.0 |

The enhancement of power density with the addition of the metallic cover, observed at gap sizes below 125 nm, can be explained by comparing the spectral heat flux absorbed by the PV cell, $q_{\omega,14} = \hbar\omega G_{\omega,14}$, for the optimized Basic and MC designs. Figure 4a shows the design parameters for the optimal configurations at a gap size of $d$ = 25 nm, while the corresponding $q_{\omega,14}$ for these two configurations are presented in Fig. 4b. The angular frequency associated with the bandgap energy of the PV cell is also specified in Fig. 4b. For both configurations, absorption below the bandgap is minimal, which demonstrates the effectiveness of the optimal designs in reducing sub-bandgap photon absorption. The heat flux spectra have two distinct peaks: one below and one above the bandgap frequency. It is seen from Fig. 4b that the peak above the bandgap, which contributes to electron-hole pair generation, is approximately four times higher in the MC configuration than in the Basic setup, resulting in significantly greater power density. To understand the origin of these two peaks and the reason underlying the enhancement of the above-bandgap peak for the MC configuration, we analyze the modal distribution of the absorbed heat flux in the PV cell. The heat flux in the PV cell (medium 4) due to the absorption of a mode $(\omega, k_\rho)$ radiated by the emitter (medium 1) is given by $q_{\omega,k_\rho,14} = \hbar\omega \left[F_{\omega,k_\rho,14}(z_3) - F_{\omega,k_\rho,14}(z_4)\right]$, where $F_{\omega,k_\rho,14}$ is found using Eq. 5. Figures 4c and 4d show $q_{\omega,k_\rho,14}$ for the Basic and MC configurations, respectively, where the wavevector $k_\rho$ is normalized by the magnitude of the wavevector in the free space, $k_0$, in these figures. Dispersion lines for the relevant surface modes are also shown and labeled in the modal distributions presented in Figs. 4c and 4d.

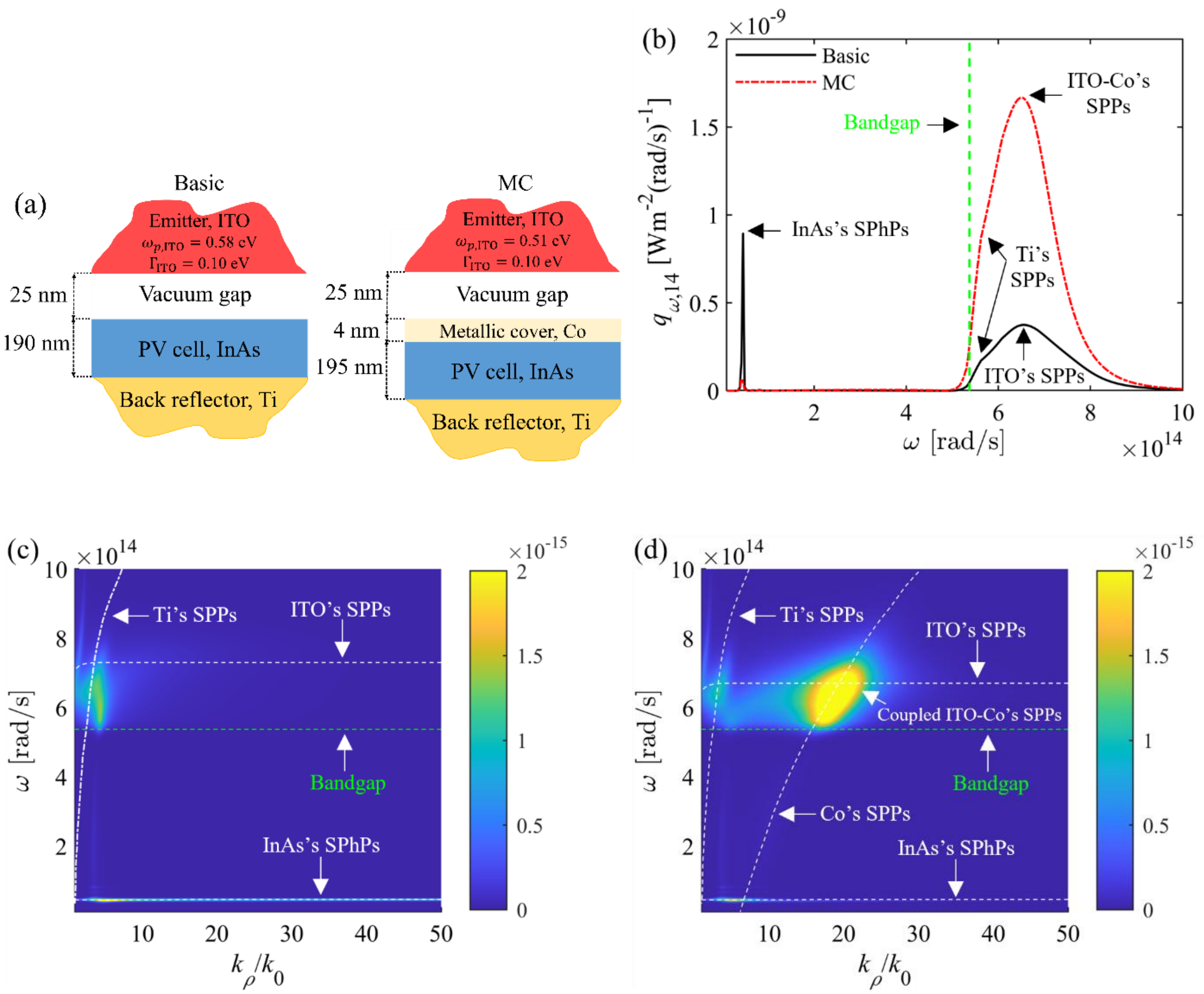

Fig. 4. (a) Schematics of nanogap TPVs with Basic and MC configurations, optimized for maximum power density. The vacuum gap size is 25 nm. (b) Spectral heat flux absorbed by the PV cell for the devices shown in Panel a. (c, d) Modal distribution of absorbed heat flux in the PV cell, $q_{\omega,k_\rho,14}$, for the (c) Basic and (d) MC configurations shown in Panel a. The corresponding dispersion relations are also plotted in Panels c and d, and the unit of the color bars in these panels is Wm$^{-2}$(rad/s)$^{-1}$m.

Three dispersion lines are present in the modal distribution of the heat flux for the Basic configuration (Fig. 4c), which are associated with the SPPs of the ITO emitter, SPhPs of the InAs PV cell, and SPPs of the Ti back reflector. The MC configuration (Fig. 4d) exhibits an additional dispersion line with a positive slope which corresponds to SPPs of the Co metallic cover excited at the Co-vacuum and Co-InAs interfaces and coupled inside this thin metallic layer.

By comparing the absorbed heat flux spectra in Fig. 4b with the modal distributions in Figs. 4c and 4d, we conclude that the sub-bandgap peak arises from the SPhPs of the InAs cell. The above-bandgap peak, which dominates the absorbed heat flux, is due to the SPPs of the ITO emitter. The modal distribution for the MC configuration (Fig. 4d) has a region of highly contributing modes around this peak frequency, which is absent for the Basic case. These modes, which are located at the intersection of the dispersion relations of the SPPs of the ITO and metallic cover, are due to the strong SPPs coupling across the vacuum gap. This coupling resonantly excites SPP modes at the interfaces of the metallic cover, which can then penetrate into and be absorbed by the InAs cell. The absorption of the coupled SPPs in the PV cell significantly increases the heat flux around the ITO SPP frequency, and thus the power density, for the MC configuration. It should be mentioned that the SPPs of the Ti back reflector also introduce a small shoulder in the absorbed heat flux spectra at a frequency of $5.6\times10^{14}$ rad/s for both Basic and MC configurations, as seen in Fig. 4b.

However, as can be seen from Fig. 3a, the effect of the metal cover on the power density vanishes for vacuum gaps larger than 125 nm. As previously discussed, the enhancement in power density results from the coupling of SPPs across the vacuum gap. These modes are evanescent within the gap and decay exponentially as $e^{ik_{0z}z}$, where $k_{0z} = \sqrt{k_0^2 - k_\rho^2}$ is the component of the free-space wavevector perpendicular to the interface, and $z$ is the distance from the interface. Since SPPs are characterized by large parallel components of the wavevector ($k_\rho \gg k_0$), $k_{0z} \approx ik_\rho$ and thus $e^{ik_{0z}z} \approx e^{-k_\rho z}$ [50,51]. This shows that SPPs decay rapidly away from the interface and cannot effectively couple across the vacuum gap when the gap size is large. As a result, the presence of the metallic cover has no impact on the power density when $d \geq 125$ nm.

The efficiency for the optimal designs found for maximal power density is shown in Fig. 3b. This figure shows that the device efficiency drops in the presence of the metallic cover. This reduction happens because, despite its minimal thickness, the metallic cover parasitically absorbs a significant portion of the thermal radiation from the emitter. However, this absorbed emission in the metallic cover does not contribute to electron-hole pair generation, and only increases the net radiative heat flux across the vacuum, $P_{rad}$, given by Eq. 12. Although the metallic cover increases the power density, $P$, the enhancement of the net radiative heat flux $P_{rad}$ is more significant, leading to a reduction in efficiency defined as $\eta = P/P_{rad}$. Nevertheless, employing a metallic cover can be beneficial for waste heat recovery applications, where power density is of more significance than efficiency.

In summary, Figs. 3a and 3b show that when maximum power density is desired, the MC configuration is the optimal structure for gaps smaller than 125 nm. In contrast, the Basic configuration generates the highest power density and provides large efficiency for larger gaps.

The optimal emitter material highly depends on the vacuum gap size. For small gaps ($d \leq 125$ nm), where the MC configuration yields the maximum power density, ITO is the optimal emitter. For larger gaps ($d$ > 125 nm) with the Basic configuration performing the best, SiC becomes the optimal choice. This is mainly because of the ability to tune the plasma frequency and damping rate parameter of ITO, which makes it possible to support SPPs at frequencies above the bandgap of the PV cell. At smaller gap sizes, the resonant SPP modes significantly enhance electron-hole pair generation in the PV cell, consequently enhancing the power density of the device (see Figs.

4c and 4d for the dispersion relation of the SPPs of ITO relative to the PV cell bandgap). At larger gap sizes, where the Basic configuration becomes optimal, SiC outperforms ITO as the emitter. This shift can be understood by examining Fig. 5b, which compares the heat flux absorbed by the PV cell for the optimal device at 200 nm employing a SiC emitter with another one optimized for the same gap size but using an ITO emitter. Both devices utilize the Basic configuration. The schematic representations of these devices are shown in Fig. 5a. The modal distributions of the absorbed heat flux for the two devices employing ITO and SiC emitters are also plotted in Figs. 5c and 5d. The spectrum of the absorbed heat flux for the ITO emitter shows a peak above the bandgap frequency. As seen from Fig. 5c, this peak is due to the excitation of ITO's SPPs. However, the contribution of this peak to the total absorbed heat flux is not as dominant as that for the smaller gaps. This reduced contribution is due to the fact that only surface modes with a small wavevector ($k_\rho < 2k_0$) can tunnel a relatively large gap of size 200 nm. On the other hand, ITO has a smaller refractive index at frequencies greater than the bandgap of the PV cell (~ 0.6 – 1.5) than SiC (~ 2.6). As such, ITO is less effective in supporting FTIR modes, which have a larger decay length than surface modes, than SiC. As such and as seen from Fig. 5b, the heat flux at above-bandgap frequencies drops sharply for the ITO emitter beyond its SPP frequency. In contrast, SiC supports FTIR modes over a wide spectral range above the bandgap frequency, leading to more broadband absorption and thus higher power generation.

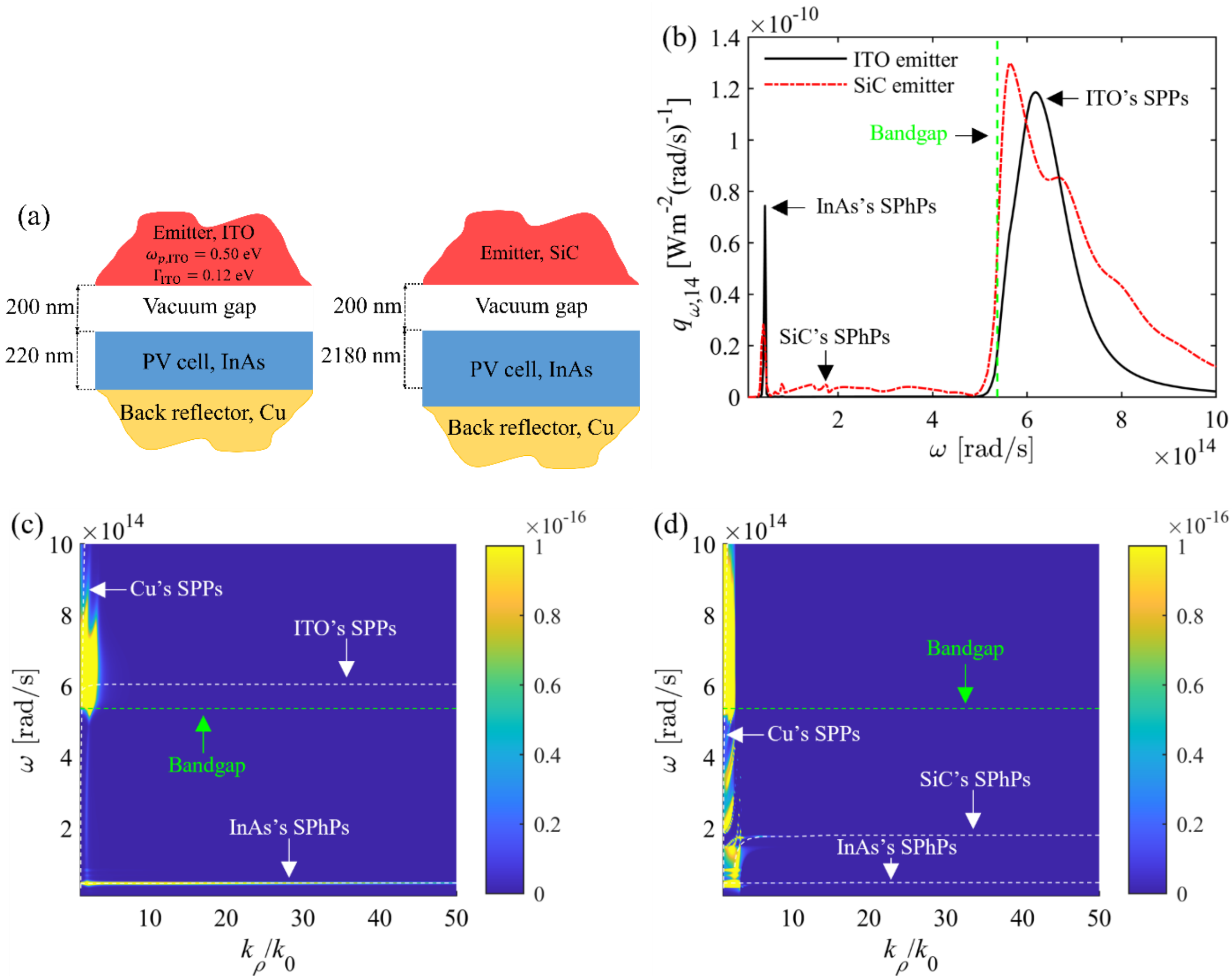

Fig. 5. (a) Schematics of the nanogap TPV devices optimized for maximal power density at a gap size of $d$ = 200 nm. The device on the left is constrained to an ITO emitter, while the device on the right utilizes a SiC emitter. (b) Spectral heat flux absorbed by the PV cell, $q_{\omega,14}$ , for the two devices shown in Panel a. (c, d) The modal distribution of absorbed heat flux by the PV cell, $q_{\omega,k_\rho,14}$ , for the devices utilizing (c) ITO and (d) SiC emitter.

Table 2 shows that the optimal material for the metallic cover also depends on the gap size. For the smallest gap considered ($d$ = 25 nm), Co is the most effective material. However, by increasing the gap beyond 25 nm, Cu becomes the preferred choice. This trend can be explained by considering two factors. First, the dispersion curves of SPPs for thin metallic films have a positive slope and do not flatten at large wavevectors (see, for example, the Co curve in Fig. 4d). This slope depends on the plasma frequency of the metal, with higher plasma frequencies resulting in steeper dispersion curves. Therefore, as can be seen from Figs. 4d and 5c, the dispersion curve for SPPs of Co with a plasma frequency of 4.0 eV [52] is less steep than that of Cu with a plasma frequency of 7.4 eV [52]. Second, the $k_\rho$ values of the SPP modes contributing to the heat flux decrease as the gap size increases, which is due to the exponential decay of their magnitude with distance as $e^{-k_\rho z}$. Since the contributing SPPs are located at smaller $k_\rho$s for larger gaps, efficient heat transfer requires the SPPs of the emitter and the metallic cover to couple at lower $k_\rho$. This is more readily achieved with Cu, which has a higher plasma frequency and thus a steeper dispersion curve than Co. At smaller gaps, however, Co is preferred as its lower SPP dispersion slope enables coupling at large wavevectors.

Table 2 shows that the optimal thickness of the metal layer is very small, ranging from 3 to 4 nm. The reason is that metals are highly absorbing, and a thicker metal layer can absorb a significant portion of the in-band photons radiated from the emitter before they reach the PV cell. An ultrathin metal layer is ideal as it enables SPP coupling across the vacuum gap while minimally absorbing thermal radiation coming from the emitter.

Table 2 also shows that InAs outperforms GaSb as the PV cell for all vacuum gap sizes. The reason is the smaller bandgap of InAs (0.354 eV) compared to GaSb (0.726 eV). At an emitter temperature of 900 K, a great portion of the emitted photons have energies between the bandgaps of InAs and GaSb. While these photons are out-of-band for GaSb, InAs can utilize them as in-band photons. This allows InAs to absorb a greater portion of the emitter's radiation and consequently generate more power.

It is observed from Table 2 that the choice of back reflector material is also gap-dependent. Ti is preferred for vacuum gaps up to 100 nm, while Cu becomes more suitable for larger gaps. The back reflector should have a low damping rate to minimally absorb photons. Among the metals considered for the back reflector, Cu, which is the optimal choice at large gaps, has the smallest damping rate, as shown in Table S-1 of the supplementary materials. For smaller gaps, where surface modes can couple efficiently across the vacuum, Ti is preferred due to its lower plasma frequency than Cu (2.5 versus 4.0 eV [52]) and moderate damping rate. As shown in Fig. 4d, the small slope of Ti's dispersion curve enables coupling of its SPPs with those of ITO. This SPPs coupling increases radiative heat transfer to the PV cell, thus enhancing the device's power density.

### 4.2. Optimal designs for maximum efficiency

Figure 6a shows the maximum efficiency achieved for each of the four configurations as a function of gap size, while the corresponding power densities for these optimal cases are presented in Fig. 6b. Table S-4 in the supplementary materials lists the optimal design variables and performance metrics for each configuration and vacuum gap size. A summary of these results is provided in Table 3, which shows the configuration with the highest efficiency along with its associated design variables and performance parameters as a function of gap size. Similar to the trend previously observed from Fig. 3b, Fig. 6a shows that utilizing a metallic cover for the PV cell reduces the efficiency of the device. This reduction is due to the substantial absorption of photons in the metallic cover, which increases $P_{\text{rad}}$ given by Eq. 12. As such, when maximizing efficiency is the primary design objective for the nanogap TPV device, using a metallic cover is not recommended. However, as discussed in Section 4.1, a metallic cover can enhance power density for gap sizes smaller than 125 nm, making it beneficial when maximal power density is desired.

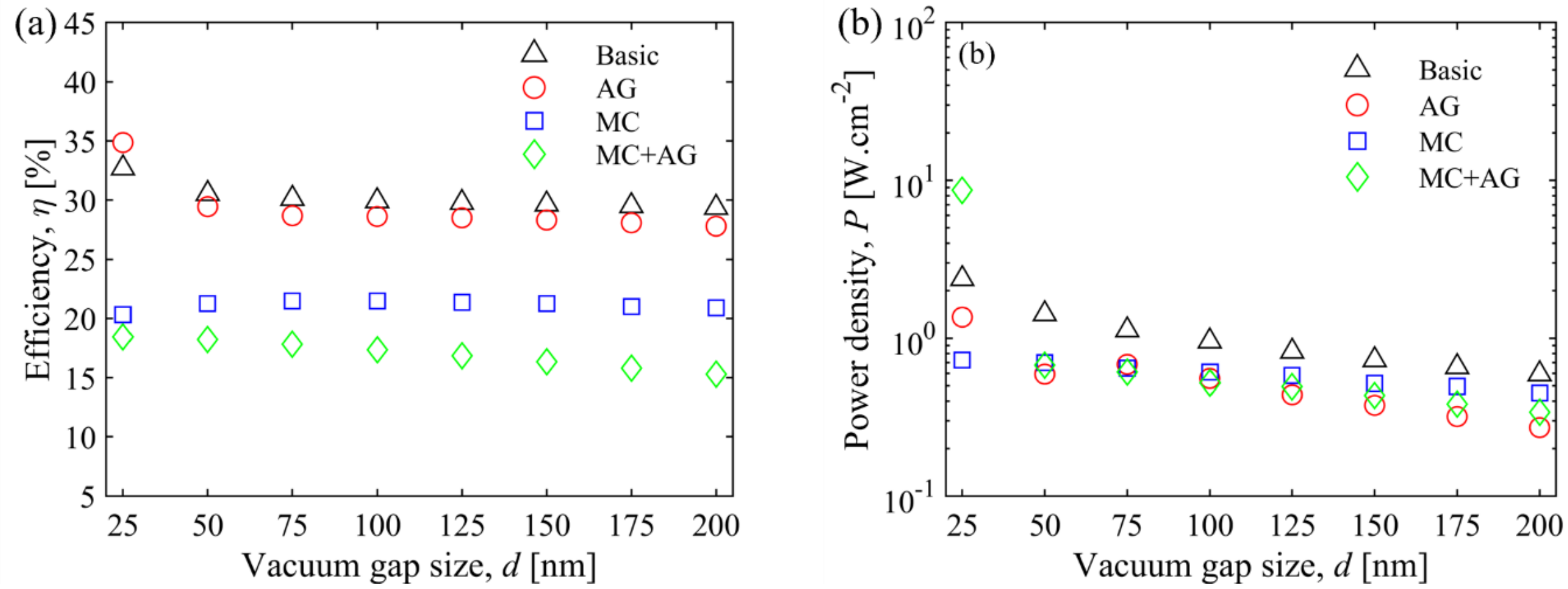

Fig. 6. (a) Efficiency and (b) power density of nanogap TPV devices optimized for maximal efficiency as a function of vacuum gap size.

Table 3. Optimal design variables and performance parameters for maximal *efficiency*.

| $d$ [nm] | Config. | $M_e$ | $\omega_{p,\mathrm{ITO}}$ [eV] | $\Gamma_{\mathrm{ITO}}$ [eV] | $M_m$ | $t_m$ [nm] | $M_{PV}$ | $t_{PV}$ [nm] | $M_b$ | $t_a$ [nm] | $M_{br}$ | $P$ [Wcm$^{-2}$] | $\eta$ [%] |
|---|---|---|---|---|---|---|---|---|---|---|---|---|---|
| 25 | AG | ITO | 0.51 | 0.1 | - | 0 | InAs | 40 | GaP | 1270 | Cu | 1.36 | 34.9 |
| 50 | Basic | ITO | 0.55 | 0.1 | - | 0 | InAs | 275 | - | 0 | Cu | 1.42 | 30.5 |
| 75 | Basic | ITO | 0.53 | 0.1 | - | 0 | InAs | 260 | - | 0 | Cu | 1.12 | 30.1 |
| 100 | Basic | ITO | 0.52 | 0.1 | - | 0 | InAs | 250 | - | 0 | Cu | 0.95 | 29.9 |
| 125 | Basic | ITO | 0.50 | 0.1 | - | 0 | InAs | 240 | - | 0 | Cu | 0.82 | 29.8 |
| 150 | Basic | ITO | 0.49 | 0.1 | - | 0 | InAs | 235 | - | 0 | Cu | 0.72 | 29.6 |
| 175 | Basic | ITO | 0.49 | 0.1 | - | 0 | InAs | 225 | - | 0 | Cu | 0.65 | 29.5 |
| 200 | Basic | ITO | 0.48 | 0.1 | - | 0 | InAs | 220 | - | 0 | Cu | 0.59 | 29.4 |

Figure 6a shows that the optimal Basic and AG configurations, which do not utilize a metallic cover, have the highest efficiencies. The efficiencies of the AG and Basic configurations are close. The AG configuration has a slightly larger efficiency at a vacuum gap size of $d$ = 25 nm, while the Basic configuration is marginally more efficient at gaps greater than 25 nm. However, for all considered gaps, the Basic configuration generates a significantly larger power density than the AG configuration (1.6 to 2.4 times larger).

The limited efficiency gain when using the AG configuration may seem unexpected, considering that the presence of an air gap is generally known to enhance device efficiency [38]. The AG configuration allows using thinner PV cells for the device without affecting the power density significantly. In the presence of the air gap, the photons that are not absorbed in the PV cell on the first pass can be reflected back toward the cell when they reach the cell-air interface. In this way, the reflected in-band photons find a second chance to be absorbed in the PV cell, while out-of-band photons are sent back to the emitter instead of being absorbed by the back reflector. Decreasing the thickness of the PV cell can increase the efficiency of the device by reducing the out-of-band absorption and non-radiative recombination, while increasing the open-circuit voltage. However, thin PV cells must be mechanically supported by a buffer layer, as discussed earlier. The presence of this buffer layer increases the reflectivity at the vacuum-cell interface, resulting in reduced thermal radiation flux from the emitter to the PV cell. This reduced heat transfer causes a drop in power generation, which also lowers the efficiency. To compensate for this reduced heat flux, a thicker PV cell is needed when using a buffer, which diminishes the benefits of the air gap.

For the maximal efficiency configurations listed in Table 3, ITO is found as the optimal emitter. The optimal performance of ITO is due to its tunable optical properties. ITO can be engineered to support SPPs above the PV cell's bandgap while also minimizing out-of-band absorption. This results in a large ratio of useful in-band emission to detrimental out-of-band absorption by the PV cell, leading to enhanced efficiency. As discussed in Section 4.1, a SiC emitter yields the highest power density for large gaps. However, SiC is not a very efficient emitter for large separation gaps. This is because SiC's high power density at larger gaps (e.g., $d$ = 200) is enabled by using thick PV cells (e.g., $t_{\text{PV}}$ = 2180 nm). Thick PV cells increase out-of-band photon absorption and reduce the open-circuit voltage, both of which degrade the efficiency of the device. The reduction in open-circuit voltage happens to suppress non-radiative recombination losses (i.e., Auger and SRH recombination), which are significant in thick cells. The Auger and SRH losses scale linearly with PV cell thickness, while they increase exponentially with the open-circuit voltage.

As shown in Table 3, GaP is chosen as the buffer material for the AG configuration, which is the optimal configuration for a vacuum gap of size 25 nm. GaP is more advantageous than InP and GaAs because of its wider bandgap, $\omega_g$, and lower imaginary part of the dielectric function, $\varepsilon''$, across most of the relevant spectrum (see Fig. S-1f in the supplementary materials for $\varepsilon''$ of these three semiconductors). Since semiconductors have significantly higher absorption above their bandgap energy, GaP's larger bandgap results in lower absorption over a broader spectral range. The larger $\omega_g$ and smaller $\varepsilon''$ of GaP minimizes photon absorption in the buffer layer. This minimal absorption is desired, since photons not absorbed by the PV cell should ideally pass through the buffer, reflect off the back reflector, and return to the cell for another chance at power generation.

Table 3 shows that the optimal air gap is about 1 μm. This relatively large gap between the buffer and the back reflector ensures that the evanescent modes at the buffer-air interface decay sufficiently before reaching the back reflector. In this manner, absorption at the back reflector is minimized, benefiting the device's efficiency. It should be noted that our simulations show that both power density and efficiency remain nearly constant when the air gap increases to the upper limit of 5 μm considered in this study. Therefore, a larger air gap may be utilized if it simplifies fabrication.

Additionally, Cu is selected as the optimal back reflector material for obtaining maximum efficiency for all considered gap sizes. When optimizing for efficiency, it is crucial to minimize photon absorption in all layers except for the PV cell. The photon absorption in metals is related to their damping rate. As seen in Table S-1 in the supplementary materials, Cu has the lowest damping rate and thus photon absorption among the considered metals.

### 4.3. Optimal Designs for Trade-off between Power Density and Efficiency

Sub-sections 4.1 and 4.2 analyzed the nanogap TPV configurations that maximize power density and efficiency, respectively. However, as shown in Fig. 2b, the configuration that has the highest power typically yields the lowest efficiency, while the most efficient configuration generates the smallest power density. In many cases, finding a balance between power density and efficiency is necessary, as both parameters can have equal importance depending on the desired application. This section discusses designs corresponding to the optimal trade-off between these two competing performance parameters as a function of vacuum gap size. As previously discussed, the optimal trade-off point is found using the TOPSIS method [49].

Figures 7a and 7b, respectively, present the power density and efficiency of the optimal trade-off design versus the gap size for the four considered configurations. The optimal design variables and performance parameters obtained for each configuration and gap size are listed in Table S-5 in the supplementary materials. For each gap size, the configuration with the largest closeness coefficient, determined using the TOPSIS method [49], is selected as the overall optimum. The optimal configuration along with its design variables, power density, and efficiency are presented in Table 4.

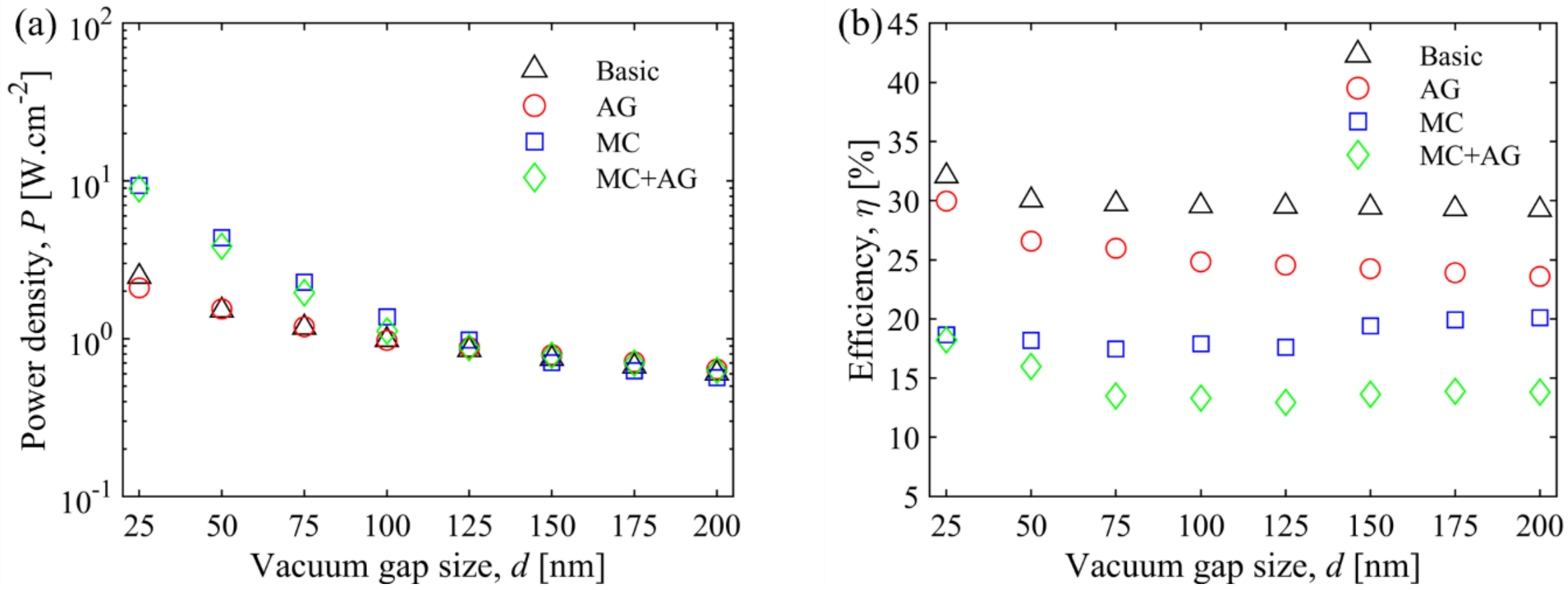

Fig. 7. (a) Power density and (b) efficiency of nanogap TPV devices optimized for a trade-off between power density and efficiency, shown as a function of vacuum gap size.

Table 4. Optimal design variables and performance parameters for a *trade-off between power density and efficiency*.

| $d$ [nm] | Config. | $M_e$ | $\omega_{p,\mathrm{ITO}}$ [eV] | $\Gamma_{\mathrm{ITO}}$ [eV] | $M_m$ | $t_m$ [nm] | $M_{\mathrm{PV}}$ | $t_{\mathrm{PV}}$ [nm] | $M_b$ | $t_a$ [nm] | $M_{\mathrm{br}}$ | $P$ [Wcm$^{-2}$] | $\eta$ [%] |
|---|---|---|---|---|---|---|---|---|---|---|---|---|---|
| 25 | MC | ITO | 0.61 | 0.11 | Cu | 3 | InAs | 340 | - | 0 | Cu | 9.34 | 18.7 |
| 50 | MC | ITO | 0.54 | 0.1 | Cu | 3 | InAs | 325 | - | 0 | Cu | 4.36 | 18.2 |
| 75 | MC | ITO | 0.53 | 0.1 | Cu | 3 | InAs | 320 | - | 0 | Cu | 2.28 | 17.5 |
| 100 | Basic | ITO | 0.55 | 0.15 | - | 0 | InAs | 245 | - | 0 | Cu | 0.98 | 29.6 |
| 125 | Basic | ITO | 0.53 | 0.14 | - | 0 | InAs | 235 | - | 0 | Cu | 0.85 | 29.5 |
| 150 | Basic | ITO | 0.52 | 0.13 | - | 0 | InAs | 230 | - | 0 | Cu | 0.74 | 29.4 |
| 175 | Basic | ITO | 0.51 | 0.13 | - | 0 | InAs | 220 | - | 0 | Cu | 0.65 | 29.3 |
| 200 | Basic | ITO | 0.50 | 0.12 | - | 0 | InAs | 220 | - | 0 | Cu | 0.60 | 29.2 |

The trends in Figs. 7a and 7b are similar to those found from Figs. 3 and 6 for the optimal designs obtained for maximal power density and maximal efficiency, respectively. Figure 7 shows that using a metal cover significantly boosts the power density for gaps smaller than 125 nm, at the cost of an appreciable drop in efficiency. It is seen from Fig. 7b that the Basic structure delivers the highest efficiency. Based on the data presented in Table 4, the MC configuration is the best trade-off option for gap sizes up to 75 nm, delivering a large power density and a moderate efficiency. For larger gaps, the Basic configuration, which has high efficiency and reasonable power density, becomes optimal. Similar to configurations optimized for maximum efficiency, ITO is found as the best-performing emitter. ITO's tunable plasma frequency and damping rate allow for maximizing the ratio of in-band to out-of-band emission. Copper is the optimal material for both the metal cover and back reflector due to its low damping rate, which minimizes photon absorption and enhances efficiency.

### 4.4. Sensitivity analysis

#### 4.4.1. Most influential parameters on the outputs

To identify the parameters with the greatest impact on the outputs, total-order Sobol indices [53-54] are computed for all design parameters. The total-order Sobol index measures the overall contribution of the parameter to the output variance, capturing both its individual effect and its interactions with other parameters [53-54]. These indices are particularly suitable for nonlinear models with multiple design parameters. The Sobol indices are bounded between 0 and 1, with larger indices indicating a stronger impact on the output. The Sobol indices are calculated for all four structures, namely, MC+AG, MC, AG, and Basic, at two gap sizes of $d$ = 25 and 100 nm. As an example, the total-order Sobol indices for the power density and efficiency of the MC+AG structure are shown in Fig. 8. The results show that, for all structures, the emitter ($M_\mathrm{e}$) and PV cell ($M_{\mathrm{PV}}$) materials are consistently among the three most influential parameters. For the Basic and AG structures, which do not employ a metallic cover, the PV cell thickness ($t_{\mathrm{PV}}$) is among the three most influential parameters. In contrast, for the MC+AG and MC structures with a metallic cover, the material for the metallic cover ($M_{\mathrm{m}}$) ranks among the three most influential parameters.

Emitter material plays an important role, as it determines the spectrum of radiative heat flux incident on the PV cell. Since only above-bandgap phonons contribute to power generation, an emitter with maximal emission above the bandgap and minimal emission below the bandgap is desired. The PV cell material is important as its bandgap determines which portions of the heat flux spectrum can generate power. Compared to GaSb, a larger fraction of the incident spectrum is located above the bandgap for InAs. The thickness of PV cell is important, as it directly affects the recombination losses as well as the in-band and out-of-band absorption of the cell. The metallic cover material strongly affects the outputs because of its parasitic absorption and SPPs dispersion characteristics. An optimal metallic cover enhances the power density through SPP coupling, while minimally absorbs the heat flux incident from the emitter.

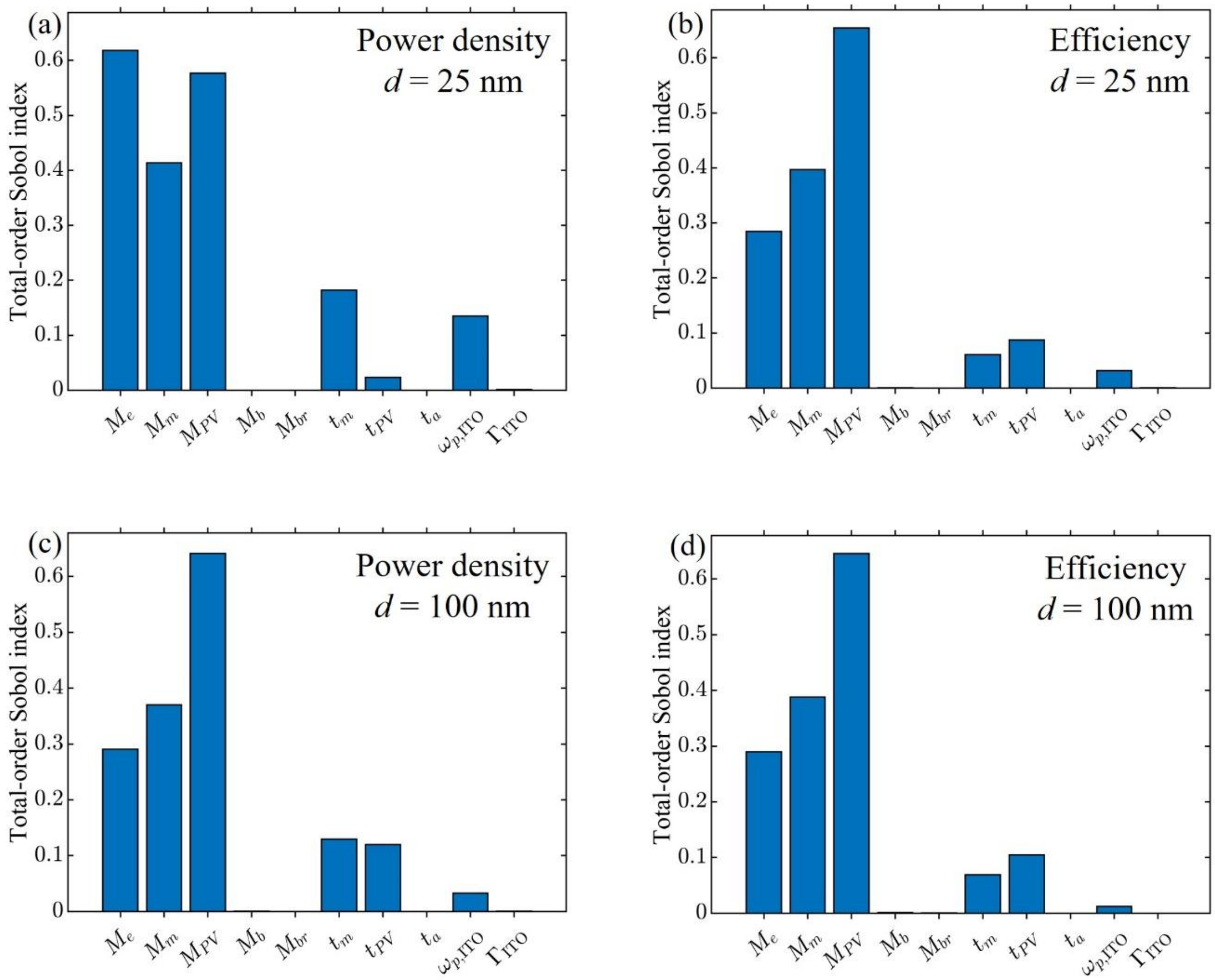

Fig. 8. Total-order Sobol indices for the MC+AG structure. Panels (a) and (b) show the indices for power density and efficiency, respectively, at a gap of $d$ = 25 nm, while Panels (c) and (d) represent the corresponding indices for $d$ = 100 nm.

**4.4.2. Effect of metallic-cover-induced nonradiative recombination loss on performance**

The addition of the metallic cover can increase the SRH recombination rate in the PV cell through introducing recombination-active metal-semiconductor interface states and modifying the near-surface carrier distribution [55]. This effect can be modeled by replacing the bulk SRH lifetime with an effective lifetime, $\tau_{\text{eff}}$, including an additional term related to the surface recombination as $\frac{1}{\tau_{\text{eff}}} \approx \frac{1}{\tau_{\text{bulk}}} + \frac{S_{\text{surf}}}{t_{\text{PV}}}$, where $S_{\text{surf}}$ is the surface recombination velocity [56-57]. It should be noted that adding the term $\frac{S_{\text{surf}}}{t_{\text{PV}}}$ to the lifetime expression applies the surface recombination penalty to the entire PV cell thickness. However, the effect of metallic cover on the lifetime is expected to be significant mostly near the metal-PV cell interface [55]. As a result, this approximation may overestimate the SRH recombination loss and the corresponding negative impact on the power density and efficiency.

We conduct a case study to evaluate the impact of enhanced SRH caused by the addition of the metallic cover on the performance of the optimized devices. For this purpose, we re-evaluate the performance of an optimal MC structure obtained for a trade-off between maximum power density and efficiency at a vacuum gap of 25 nm, while accounting for the increased SRH recombination. We consider a moderately affected PV cell with a surface recombination velocity of $S_{\text{surf}} = 5 \times 10^3$ cm/s, as well as a severely affected PV cell with $S_{\text{surf}} = 10^5$ cm/s [58-59]. For $S_{\text{surf}} = 5 \times 10^3$ cm/s, the power density and efficiency decrease by ~8% compared to the case where surface recombination is neglected. For $S_{\text{surf}} = 10^5$ cm/s, the effect is much more significant, resulting in ~50% reduction in both performance parameters. It should be noted that, as mentioned before, these reductions represent a worst-case estimate by assuming $S_{\text{surf}}$ to the entire PV cell thickness. It is also worth noting that the negative impact of the metallic cover on the SRH recombination rate and overall device performance can be mitigated by engineering the doping profile of the PV cell to create an internal electric field to make the front metal a less favorable electron pathway. This, for example, can be done by a P-type front region combined with an N-type back region [60].

The effect of the metallic cover on the Auger recombination is much less significant than the SRH recombination. The Auger recombination can be affected through the modification of the carrier concentrations $n$ and $p$ near the metal-PV cell interface. Incorporating this effect would require accounting for the spatial variations of $n$ and $p$ across the PV cell thickness, which is not feasible within the lumped model employed in this study.

#### 4.4.3. Effect of thickness-dependent dielectric function of the metallic cover on performance

Reducing the thickness of a metallic film to only a few nanometers does not significantly change its plasma frequency, which depends on the free-electron density, as long as the film is continuous [61]. However, the damping rate of thin metallic films can be larger than that of the corresponding bulk metal. Thickness-dependent data are not available for the damping rate of the metals considered in this study. Additionally, the damping rate can vary significantly depending on fabrication quality. As such, we have used the damping rates of bulk metals for the thin metallic cover.

To examine how this assumption affects the final results, we perform a case study. The MC structure for three gap sizes of 25, 50, and 75 nm optimized for a trade-off between maximal efficiency and power density are selected for this study. For these structures, for which Cu is the optimal material for the metallic cover, we re-evaluated the power density and efficiency as a function of damping rate. The considered damping rates range from the bulk value for Cu (0.0091 eV) to 10 times that value (for example, the damping rate for a Cu film with a thickness of 12.3 nm is measured to be about 6.3 times higher than the bulk value [62]). Figure 9 shows the power density and efficiency versus damping rate. The results show that the power density and efficiency drop monotonically as the damping rate increases. When the damping rate increases by 10 times, the power density drops by 52% and the efficiency decreases from ~18 to ~5% (depending on the gap size). The significant reduction in power density and efficiency is due to the larger parasitic absorption by the metallic cover with a larger damping rate.

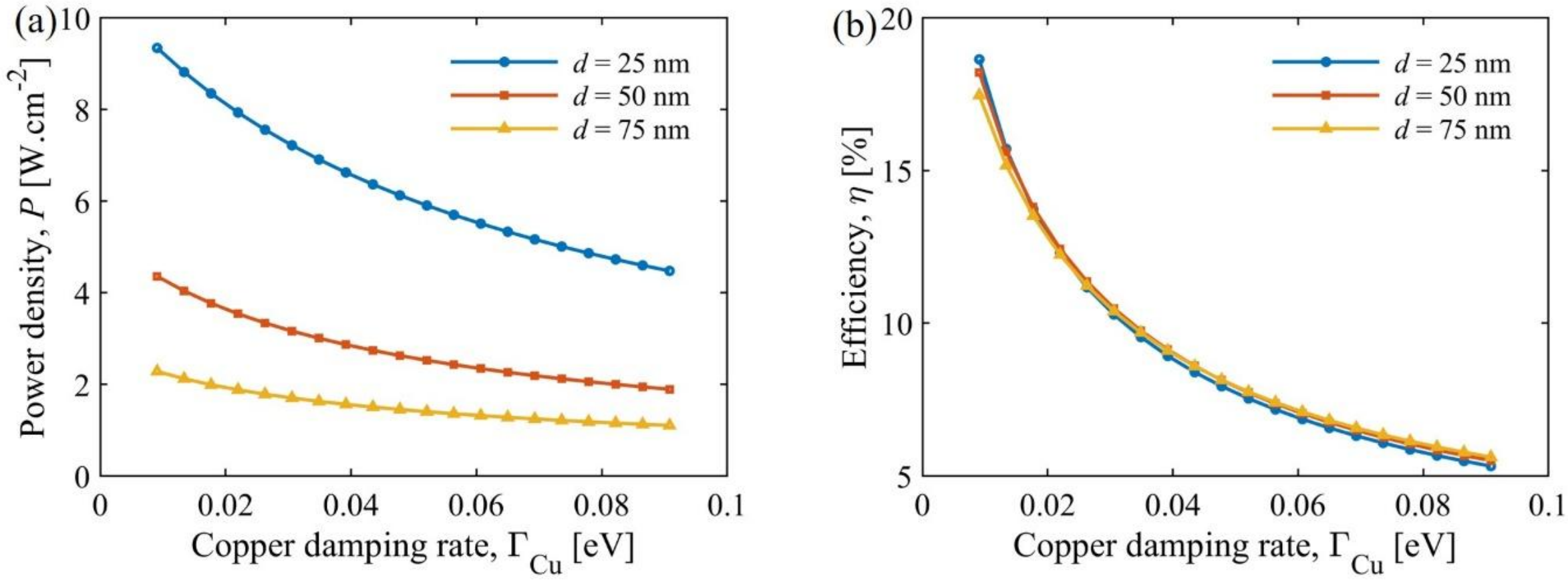

Fig. 9. (a) Power density and (b) efficiency of the MC structures optimized for a trade-off between power density and efficiency as a function of the damping rate of their copper metallic cover.

#### 4.4.4. Effect of the emitter's temperature-dependent dielectric function on performance

The emitter operates at a temperature of 900 K. The dielectric function of the emitter at this elevated temperature can be different from that at room temperature. However, available dielectric function data, particularly at the spectral range relevant to the TPV devices, is very limited. For this reason, we have used the room-temperature dielectric functions in this study.

To assess the effect of temperature dependence of the dielectric function on the performance of the optimal devices, we re-evaluated the power density and efficiency of the optimal designs with a selected SiC emitter using its dielectric function at 900 K. The dielectric function of SiC at 900 K is significantly affected around its phonon resonance frequency (~$1.8\times10^{14}$ rad/s), where the real part decreases by ~120% and the imaginary part increases by ~410%. However, the impact of this change on the device's performance is very modest. Employing the dielectric function at 900 K results in less than 1% difference in the predicted device performance for Basic and AG structures. For optimal MC structures, the change in the power density remains negligible (less than 1%), while the efficiency is affected more significantly by 10%. The power density is not significantly affected, since the resonance frequency of SiC is located below the bandgap, where absorption by the PV cell is low. The 10% reduction in efficiency for the MC structure is due to increased parasitic absorption by the metallic cover when the dielectric function at 900 K, with a larger imaginary part, is used. It should be noted that these results are specific to the SiC emitter, and extending this analysis to other emitter materials is currently not feasible due to the lack of temperature-dependent dielectric function data.

## 5. Conclusion

This study examines the performance of four various nanogap TPV configurations for recovering industrial waste heat. The considered configurations include a Basic version utilizing an emitter, a PV cell, and a back reflector, an MC configuration featuring a metallic cover on the PV cell, an AG configuration integrating an air gap between the PV cell and the back reflector, as well as an MC+AG configuration employing both the metallic cover and the air gap. It was found that the optimal configuration strongly depends on the performance objective of the device as well as the size of the vacuum gap. When maximal power output is desired, the MC configuration is the preferred choice for gaps smaller than 125 nm, as the metallic cover can enable coupling of SPPs across the vacuum gap. For gaps larger than 125 nm, the SPP coupling is not as efficient, rendering the Basic configuration optimal. It was also found that utilizing the metallic cover has a strong negative impact on the efficiency of the device due to parasitic absorption of photons in the metal. Utilizing the air gap was found to increase the efficiency for ultrathin PV cells. However, when these ultrathin PV cells are mechanically supported by a substrate, the efficiency gain diminishes. When efficiency is the primary performance objective, the Basic configuration is identified as optimal. When a tradeoff between the power density and efficiency is desired, the MC configuration is advantageous for gaps smaller than 75 nm. At larger gaps, the efficiency drop caused by the metallic cover does not justify the gain in power density, and the Basic configuration is preferred. The material choice for various components of the device has a significant effect on its performance. Among the studied materials, Cu is found as the optimal material for the metallic cover, since its SPP dispersion curve has the right slope to enable coupling of the most contributing SPP modes across the vacuum gap. InAs is identified as the optimal PV cell because of its low bandgap energy, while ITO is found as the optimal emitter due to tunability of its plasma frequency. It should be noted that while selective emitters are highly advantageous, achieving strong above-bandgap selectivity through surface resonances becomes less effective at larger, more practical gap distances, since these modes are significantly weakened. Metals with low damping rate, and thus small photon absorption, are recommended for the back reflector.

**Data availability**

The data supporting this study will be made available upon request within a reasonable time frame following publication.

**Acknowledgments**

This work used Anvil at Purdue University through allocation MCH250023 from the Advanced Cyberinfrastructure Coordination Ecosystem: Services & Support (ACCESS) program.